\documentclass{svproc}
\usepackage{url}

\usepackage{amsmath,amssymb}

\usepackage{graphicx}
\usepackage{multicol} 
\usepackage{subcaption}
\usepackage{setspace}
\usepackage[font=singlespacing]{caption}

\usepackage{color}       

\begin{document}
\mainmatter              
\title{Stochastic modeling and large-eddy simulation of heated concentric coaxial pipe flow}
%
\titlerunning{Numerical simulation of heated concentric coaxial pipe flow}  
%
\author{Marten~Klein\inst{1,2,*}
  \and Pei-Yun~Tsai\inst{1}
  \and Heiko~Schmidt\inst{1,2}
}
\authorrunning{Marten Klein, Pei-Yun Tsai, and Heiko Schmidt} 
%
%
\institute{Lehrstuhl Numerische Str\"omungs- und Gasdynamik, Brandenburgische Technische Universit\"at (BTU) Cottbus-Senftenberg, Germany\\
WWW home page: \url{https://www.b-tu.de/en/fg-stroemungsmodellierung} 
\and
Scientific Computing Lab (SCL), Energie-Innovationszentrum (EIZ), Brandenburgische Technische Universit\"at (BTU) Cottbus-Senftenberg, Germany\\
WWW home page: \url{https://www.b-tu.de/en/energie-innovationszentrum}
}

\maketitle              

\begin{abstract} 

Turbulent concentric coaxial pipe flows are numerically investigated as canonical problem addressing spanwise curvature effects on heat and momentum transfer that are encountered in various engineering applications.
It is demonstrated that the wall-adapting local eddy-viscosity (WALE) model within a large-eddy simulation (LES) framework, without model parameter recalibration, has limited predictive capabilities as signalized by poor representation of wall curvature effects and notable grid dependence.
The identified lack in the modeling of radial transport processes is therefore addressed here by utilizing a stochastic one-dimensional turbulence (ODT) model.
A standalone ODT formulation for cylindrical geometry is used in order to asses to which extent the predictability can be expected to improve by utilizing an advanced wall-modeling modeling strategy.
It is shown that ODT is capable of capturing spanwise curvature and finite Reynolds number effects for fixed adjustable ODT model parameters. 
Based on the analogy of heat and mass transfer, present results yield new opportunities for modeling turbulent transfer process in chemical, process, and thermal engineering.

\end{abstract}


\vspace*{-2ex}
\small
\subsection*{Bibliographic information on the Version of Record}
\noindent
This is a preprint of the following chapter: \\[1ex] 
Marten Klein, Pei-Yun Tsai, and Heiko Schmidt, \textit{Stochastic modeling and large-eddy simulation of heated concentric coaxial pipes}, published in \textit{New Results in Numerical and Experimental Fluid Mechanics XIV -- Contributions to the 23rd STAB/DGLR Symposium Berlin, Germany, 2022}, edited by Andreas Dillmann, Gerd Heller, Ewald Krämer, Claus Wagner, and Julien Weiss. \textit{Notes on Numerical Fluid Mechanics and Multidisciplinary Design}, vol.~154, 2024, Springer, Cham, reproduced with permission of Springer Nature Switzerland AG. \\[1ex]
The final authenticated version is available online at: \\ \url{http://dx.doi.org/10.1007/978-3-031-40482-5_41}

\normalsize
\newpage

\section{\label{sec:intro} Introduction}

Robust predictions of turbulent heat, mass, and momentum transfer crucially depend on the accurate numerical modeling of wall-normal processes.
This is a standing challenge for various engineering applications encompassing internal flows (like heat exchangers~\cite{fukuda_etal:2020}, chemical reactors~\cite{balestrin_etal:2021}, or electrostatic gas-cleaning devices~\cite{Medina_etal:2022}).
The spanwise wall curvatures that are frequently encountered in these applications further complicate the numerical analysis due to nonuniversal scaling properties of radial boundary layers that govern the transfer processes.
Complications arise from variations of diffusive to curvature length scales that should ideally be treated with DNS (e.g.~\cite{bagheri_etal:2021}) 
  but also differential diffusion effects that result from finite, in particular, very small or large Prandtl and Schmidt numbers (e.g.~\cite{Klein_etal:2022}). 
These effects tend to be more severe in wide-gap configurations that are sometimes technically unavoidable, like in coaxial electrostatic precipitators \cite{Medina_etal:2022}.
The high costs resulting from the resolution requirements severely limit the applicability of DNS.
Alternative strategies for modeling and simulation are therefore required.

A promising approach in this regard is probabilistic wall-modeled LES that utilizes stochastic turbulence modeling only for the near-wall LES cells (e.g.~\cite{Schmidt_etal:2003,Freire_Chamecki:2021}).
Another approach is ODTLES, which is a more comprehensive subgrid-scale modeling strategy, in which the 3-D flow domain is discretized with domain-spanning ODT lines that are coupled by the large-scale flow (e.g.~\cite{Gonzalez-Juez_etal:2011,Glawe_etal:2018}).
In these approaches, the so-called one-dimensional turbulence (ODT) model \cite{Kerstein:1999} is utilized in order to economically resolve all relevant scales of the flow but only along a single physical coordinate.
It has been demonstrated recently that ODT is capable of capturing differential diffusion effects in channels~\cite{Klein_etal:2022} and wall-curvature effects in heated pipes~\cite{Medina_etal:2019}.

\begin{figure}[t]
  \centering
  \includegraphics[scale=0.3]{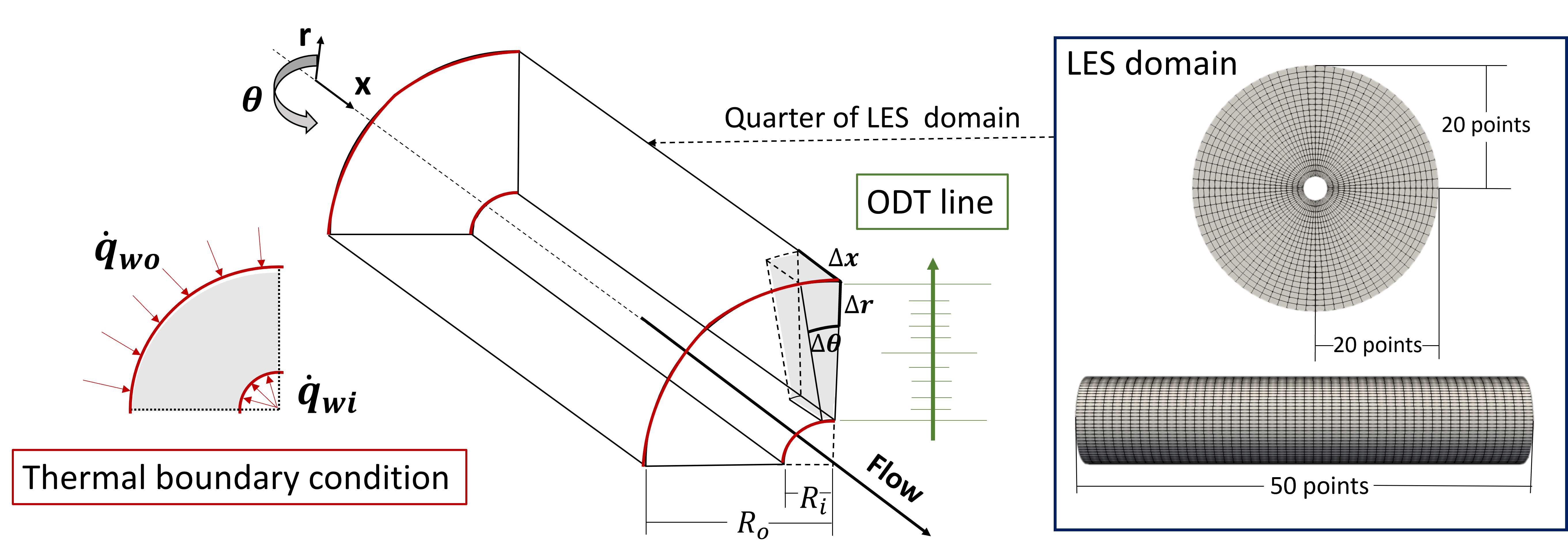} 
  \caption{%
    Schematic of the heated concentric coaxial pipe flow configuration investigated.
    Only one quarter of the pipe is shown.
    The computational domains utilized for LES (\textit{blue}) and standalone stochastic ODT simulations (\textit{green}) are highlighted by colored boxes. 
    The flow is heated by a prescribed constant heat fluxes $\dot{q}_\text{w,i}$ and $\dot{q}_\text{w,o}$ at the inner and outer cylinder wall, respectively (\textit{red}).}
  \label{fig:config}
\end{figure}

Here, we extend the previous studies by applying ODT as standalone tool to pressure-driven concentric coaxial pipe flow of notionally infinite length that is radially bounded by a heated inner (convex) and heated outer (concave) cylinder.
This serves as canonical problem for applications in heat exchangers and chemical reactors.
Figure~\ref{fig:config} shows a sketch of the set-up.
Numerical efficiency is addressed by utilizing a dynamically-adaptive cylindrical ODT formulation \cite{Lignell_etal:2018} in which the ODT domain represents an infinitesimal wedge that spans the radial gap. 

The rest of this paper is organized as follows.
In section~\ref{sec:method} we give an overview of the LES and ODT model formulations and their application to coaxial pipe flow.
In section~\ref{sec:results} we discuss the predictive capabilities of both approaches by comparing ODT and LES with available reference DNS.
Last, in section~\ref{sec:conc}, we summarize and conclude our findings.

\section{\label{sec:method} Method}

This section briefly discusses the LES and ODT numerical models. We consider an incompressible constant-property Poiseuille flow confined between two coaxial concentric cylinders driven by an axial pressure gradient as sketched in Figure~\ref{fig:config}. Weak temperature fluctuations are modeled as a passive scalar.

\subsection{\label{sec:les-overview} Overview of the large-eddy simulation set-up}

Large-eddy simulation is based on the universality of small-scale turbulence \cite{smagorinsky:1963}.
The key idea of LES is to resolve nonuniversal large and meso scale flow features, while modeling the dissipative ensemble effect of small-scale turbulent eddies based on physical relations with the resolved scales.
The universality assumptions underlying such subgrid-scale (SGS) closure models are \emph{not} generally justified, for instance, when the flow over a curved instead of planar wall is considered. 
In this study, we demonstrate the lack in predictive capabilities for a wall-modeled LES of concentric coaxial pipe flow utilizing the WALE model~\cite{nicoud:1999} previously calibrated for plane channel flow.
The filtered continuity, momentum, and scalar transport equation (per unit mass) for an incompressible flow are given by 
\begin{align}
    \frac{\partial \overline{u}_i}{\partial x_i} &=  0 \;,
    \label{eq:les_mass} \\
    \frac{\partial\overline{u}_i}{\partial t} 
    + \overline{u}_j\frac{\partial\overline{u}_i}{\partial x_j} &= 
    - \frac{1}{\rho}\frac{\partial \overline{p}}{\partial x_i}
    + \frac{\partial}{\partial x_j}\left( \nu \frac{\partial\overline{u}_i}{\partial x_j} \right) 
    + \frac{\partial \tau_{ij}}{\partial x_j} 
    + \overline{f}_i \;,
    \label{eq:les_mom} \\
    \frac{\partial \overline{\Theta}}{\partial t} 
    + \overline{u}_j\frac{\partial \overline{\Theta}}{\partial x_j} &= \frac{\partial}{\partial x_j}\left( \Gamma \frac{\partial\overline{\Theta}}{\partial x_j} \right) 
    + \frac{\partial q_j}{\partial x_j} 
    + \overline{s}_\Theta \;,
    \label{eq:les_sca}
\end{align}
\noindent %
where $t$ is time, $x_i$ ($i=1,2,3$) the Cartesian coordinates of the configuration space, $\overline{u}_i$ the Cartesian components of the filtered velocity vector, $\overline{p}$ the filtered pressure, $\overline{f}_i$ the prescribed filtered momentum sources, like a prescribed streamwise mean pressure-gradient force, $\overline{f}_i=-\rho^{-1}\,({\rm d}P/{\rm d}x)\,\delta_{i1}$, where $\delta_{ij}$ ($j=1,2,3$) is the Kronecker symbol, $\overline{\Theta}$ the filtered temperature, $\overline{s}_\Theta$ the prescribed filtered heat sources, $\rho$, $\nu$, and $\Gamma$ the constant mass density, kinematic viscosity, and thermal (passive scalar) diffusivity of the fluid, respectively, $\tau_{ij}= \overline{u_iu_j}-\overline{u}_i\overline{u}_j$ the unclosed SGS stress tensor, and $q_i=\overline{u_i\Theta}-\overline{u}_i\overline{\Theta}$ the unclosed SGS kinematic heat flux (temperature flux). 
Einstein's summation convention is implied for indices that appear twice.
By applying the Boussinesq hypothesis, the SGS stress tensor and SGS heat flux are parameterized as 
\refstepcounter{equation}
\begin{equation*}
  \tau_{ij}=2\nu_\text{t} \overline{S}_{ij} \;,
  \quad
  q_{i}=\Gamma_\text{t} \, \frac{\partial \overline{\Theta}}{\partial x_i} \;,
  \quad
  \Gamma_\text{t} = \frac{\nu_\text{t}}{Pr_\text{t}} \;,
  \eqno{(\theequation{\mathit{a},\mathit{b},\mathit{c}})}
  \label{eq:sgs}
\end{equation*}
where $\overline{S}_{ij}=(\partial\overline{u}_i/\partial x_j+\partial\overline{u}_j/\partial x_i)/2$ denotes the resolved-scale rate-of-strain tensor, $\nu_\text{t}$ the turbulent eddy viscosity, $\Gamma_\text{t}$ the turbulent thermal diffusivity, and $Pr_\text{t}=\nu_\text{t}/\Gamma_\text{t}\lesssim O(1)$ the turbulent Prandtl number, which needs to be determined empirically.
In the WALE model, $\nu_\text{t}$ is formulated by taking into account the strain rate and rotation effects, and $Pr_\text{t}$ is an empirical model constant \cite{nicoud:1999}. 

For the present study, we use the open-source CFD software \textit{OpenFOAM} \cite{Weller_etal:1998} (version~8) performing LES by adjusting the LES channel flow tutorial case utilizing the PIMPLE solver. 
As sketched in Figure~\ref{fig:config}, the LES domain is made of four identical blocks, each representing a quarter of the pipe.
The pipe length was chosen as ten times the outer radius $R_\text{o}$, that is $L=10R_\text{o}$, which means more than ten gap widths $\delta=R_\text{o}-R_\text{i}$, that is, $L/\delta>10$, where the exact value depends on the selected radius ratio, $\eta=R_\text{i}/R_\text{o}$.


\subsection{\label{sec:odt-overview} Overview of the one-dimensional turbulence model}

Kerstein's \cite{Kerstein:1999} one-dimensional turbulence model aims to resolve all relevant scales of turbulent flow. This is made feasible for high Reynolds number flows by modeling the effects of turbulent eddies by a stochastically sampled sequence of mapping events that punctuate the deterministic advancement in the form of diffusion equations.
The model aims to resolve molecular-diffusive transport processes and distinguishes them from turbulent-advective ones along the one-dimensional domain (`ODT line' in Figure~\ref{fig:config}). 
Here we limit our attention to constant-density flow treating weak temperature fluctuations as passive scalar in analogy to \cite{Klein_etal:2022} but for cylindrical geometry \cite{Lignell_etal:2018}.
The dimensionally reduced stochastic equations that are numerically solved in ODT are thus given by
\begin{align}
    \frac{\partial \tilde{u}_i}{\partial t} 
    + \sum_{t_\text{e}}\mathcal{E}_{i}\left(\tilde{u}_x,\tilde{u}_r,\tilde{u}_\theta\right)\,\tilde{\delta}\left(t-t_\text{e}\right)  &=  
    \frac{1}{r}\frac{\partial}{\partial r}\bigg(r\nu\frac{\partial \tilde{u}_i}{\partial r}\bigg)
    -\frac{1}{\rho}\frac{{\rm d}P}{{\rm d}x} \delta_{1i}
    \;,
    \label{eq:gov_vel} \\
    \frac{\partial \tilde{\Theta}}{\partial t}
    + \sum_{t_\text{e}}\mathcal{E}_{\Theta}\left(\tilde{u}_x,\tilde{u}_r,\tilde{u}_\theta\right)\,\tilde{\delta}\left(t-t_\text{e}\right) &= 
    \frac{1}{r}\frac{\partial}{\partial r}\bigg(r\Gamma\frac{\partial \tilde{\Theta}}{\partial r}\bigg)
    + \beta \tilde{u}_x 
    \;,
    \label{eq:gov_scal}
\end{align}
where $t_\text{e}$ denotes the stochastically sampled eddy occurrences, $\tilde{\delta}$ the Dirac distribution function, representing a discrete `delta kick' of infinite rate that becomes a finite `Heaviside jump' contribution upon time integration, $r$ the radial coordinate, $\tilde{u}_i(r,t)$ the model-resolved velocity vector with cylindrical components $\tilde{u}_x,\tilde{u}_r,\tilde{u}_\theta$ in axial, radial, and azimuthal direction, respectively, $-\rho^{-1}\,\left({\rm d}P/{\rm d}x\right)\, \delta_{1i}=f_i(t)$ the prescribed spatially uniform mean pressure-gradient force in axial direction, and $\beta \tilde{u}_x(r,t)=s_\Theta(r,t)$ the momentary heat sources for the model-resolved perturbation temperature $\tilde{\Theta}(r,t)$.
Following the modeling of weakly heated channel flow \cite{Klein_etal:2022}, we define $\tilde{\Theta}(r,t)=\beta x + T_0(r) - T(x,r,t)$, where $T(x,r,t)$ is an axial-radial slice of the actual temperature field at a given time, $\beta={\rm d}T_\text{m}/{\rm d}x$ the constant axial gradient of the bulk mean temperature, and $T_0(r)$ a background diffusive solution.
In the fully developed, statistically stationary state, $T_0(r)$ absorbs the physical wall temperature increase in downstream ($x$) direction so that homogeneous Dirichlet boundary conditions are prescribed for the perturbation field $\tilde{\Theta}(r,t)$. 
In available reference DNS~\cite{bagheri_etal:2021}, the wall heat fluxes are the same ($\dot{q}_\text{w,i}=\dot{q}_\text{w,o}$), whereas the wall heating rates are balanced ($\dot{q}_\text{w,i}R_\text{i}=\dot{q}_\text{w,o}R_\text{o}$) in the present ODT formulation for practical reasons.
Since boundary conditions are compensated by $T_0(r)$, no sensible difference is expected for the leading-order contribution to the perturbation temperature $\Theta$.

Note that, in contrast to LES, there is no closure and no closure modeling in ODT-
The advection and fluctuating pressure gradient terms have been formally replaced by eddy events, symbolically represented as $\mathcal{E}_i$ for the velocity and $\mathcal{E}_\Theta$ for the temperature equation. $\mathcal{E}_i$ accounts for pressure--velocity couplings that are absent in $\mathcal{E}_\Theta$. The arguments indicate that the eddy event selection is governed solely by the momentary velocity vector \cite{Klein_etal:2022}. 
Further details of the model formulation are given in \cite{Lignell_etal:2018}. 

\section{\label{sec:results} Results and discussion}

This section presents the results of the momentum and scalar transfer. 
We utilize the ODT model calibration based on heat transfer in concentric coaxial pipe flow \cite{Tsai_etal:2022}, but keep the adjustable model parameters fixed as for the LES.

For the assessment of boundary layer similarity, the friction velocity and friction temperature are defined as 
\refstepcounter{equation}
\begin{equation*}
  u_\tau = \sqrt{ \nu\left|\frac{\mathrm{d}u}{\mathrm{d}r}\right|_\mathrm{wall} } \;,
  \quad
  \Theta_\tau = \frac{\Gamma}{u_\tau}\left|\frac{\mathrm{d}\Theta}{\mathrm{d}r}\right|_\mathrm{wall} \;,
  \eqno{(\theequation{\mathit{a},\mathit{b}})}
  \label{eq:nondim}
\end{equation*}
where $u_\tau$ denotes the friction velocity and $\Theta_\tau$ the friction temperature, $u(r)=\langle\tilde{u}_x(r,t)\rangle_t$ and $\Theta(r)=\langle\tilde{\Theta}(r,t)\rangle_t$ denote temporal averaged variables in the case of ODT, but temporal, axial, and azimuthal averaged variables $u(r)=\langle\overline{u}_x(x,r,\theta,t)\rangle_{t,x,\theta}$ and $\Theta(r)=\langle\overline{\Theta}(x,r,\theta,t)\rangle_{t,x,\theta}$ in the case of LES after transformation from prognostic Cartesian to diagnostic cylindrical coordinates.

The scaled (superscript `$+$') variables are given by the normalized mean axial velocity $u^+=u/u_\tau$ and the normalized mean perturbation temperature $\Theta^+=\Theta/\Theta_\tau$ that are evaluated over the inner and outer wall for the boundary layer coordinate $r^+=|r-R_\text{i/o}|\,u_\tau/\nu$. 
The similarity parameters of the flow are given by the friction Reynolds number $Re_\tau=\delta u_\tau/\nu$ and $Pr=\nu/\Gamma=0.71$ (air). 

\subsection{\label{sec:vis} Flow visualization and computational resource requirements}

\begin{figure}[t]
    \centering
        \includegraphics[scale=0.25]{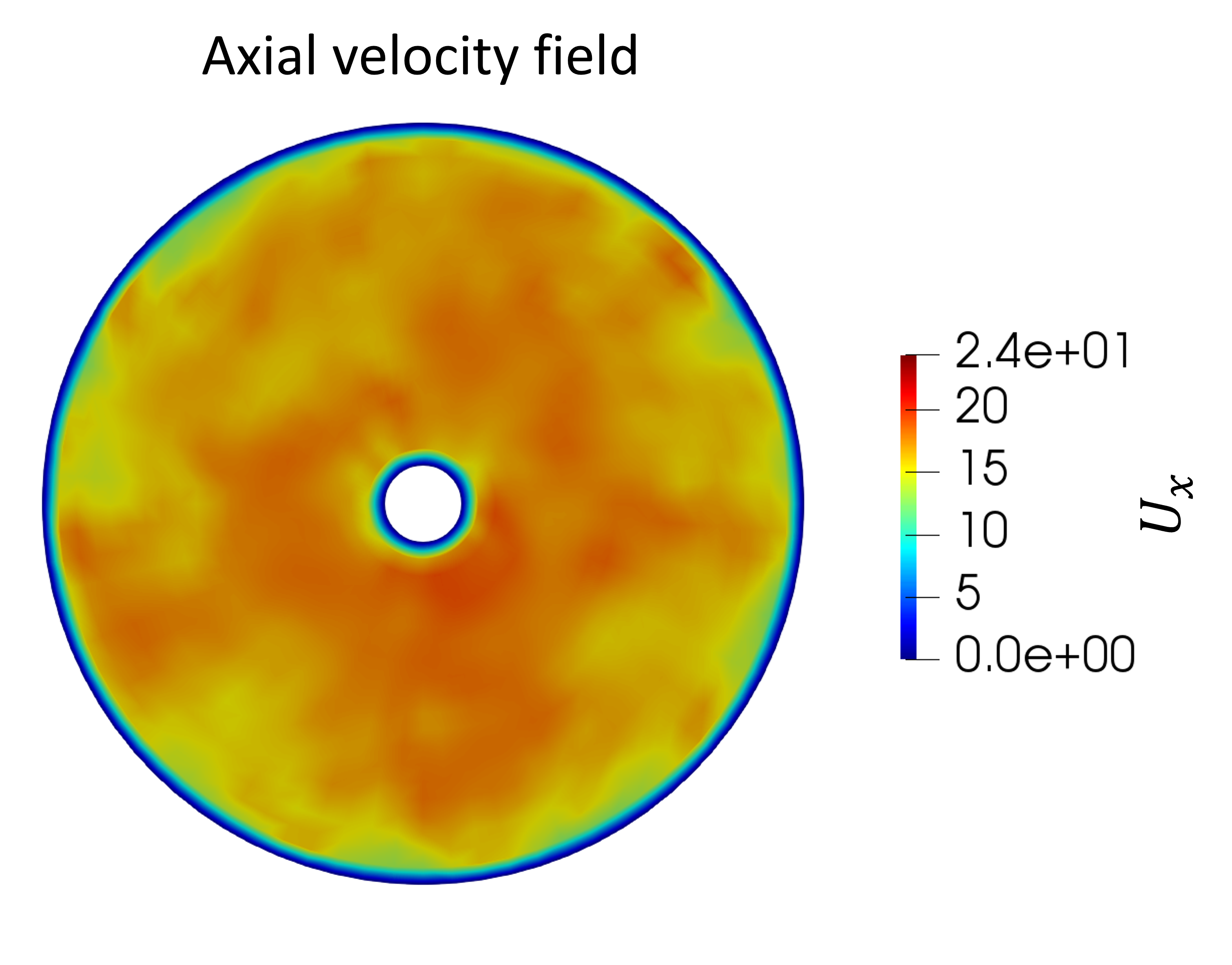}
        \includegraphics[scale=0.25]{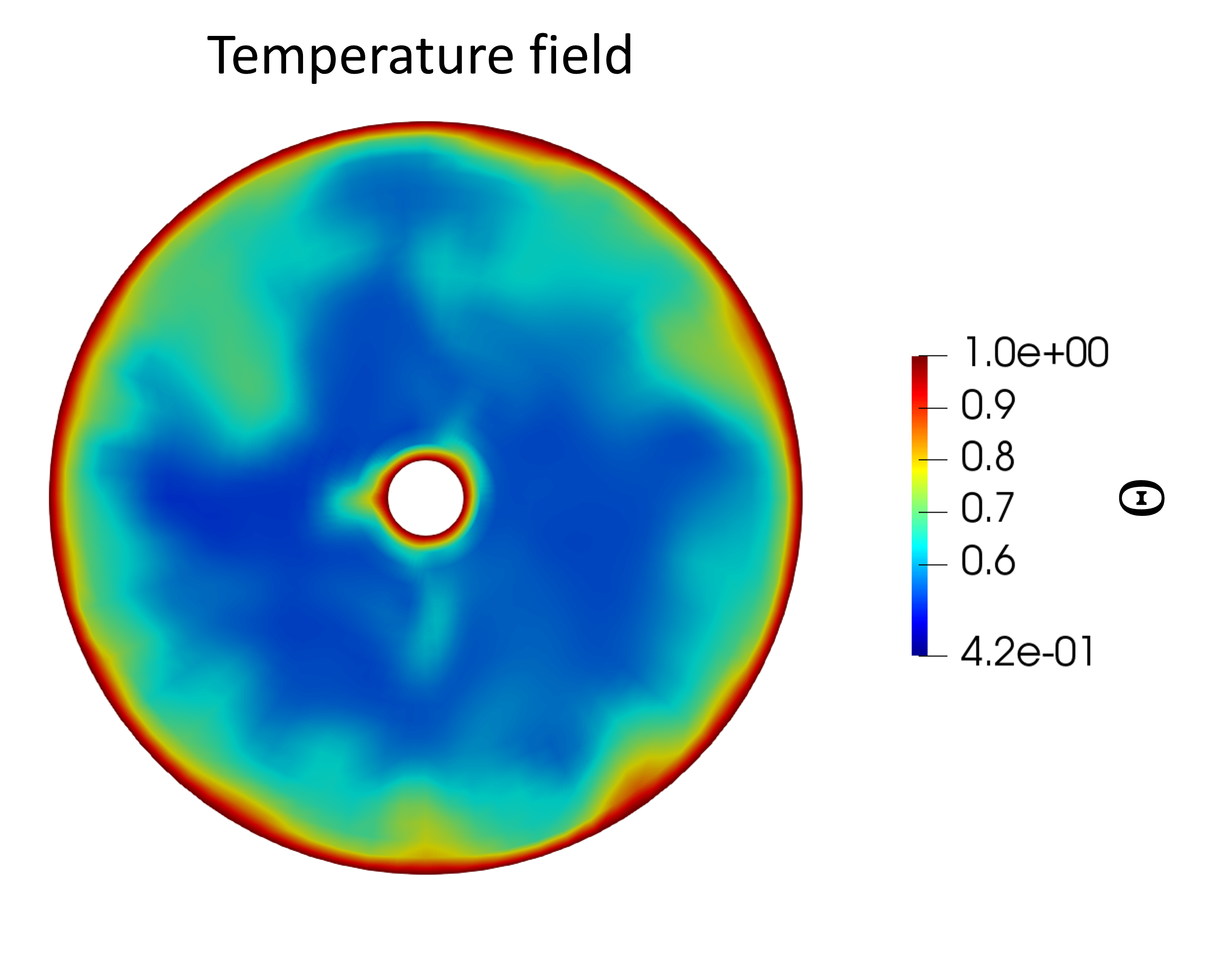}
    \caption{Contours of the (\textit{left}) momentary axial velocity and (\textit{right}) momentary perturbation temperature in a radial-azimuthal section of the LES results for the periodic concentric coaxial pipe with $\eta=0.1$ at $Pr=0.71$ and equivalent circular pipe friction Reynolds number $Re_* = 2U_* R_\text{o}/\nu = 2780$, where $U_*=\sqrt{-R_\text{o}/(2\rho)\,(\text{d}P/\text{d}x)}$. 
    }
    \label{fig:anima}
\end{figure}

Figure \ref{fig:anima} shows an radial-azimuthal section of the momentary LES solution from the statistically stationary state at $x/L=0.5$ for a representative case.
The flow is visualized by contours of the axial velocity component $\overline{u}_x$ and the temperature scalar $\overline{\Theta}$.
The results demonstrate that the region over the strongly curved inner cylinder exhibits small scale turbulence, making the case numerically more challenging then circular pipe flow.

We note that the LES for $\eta=0.1$ with a resolution of $N_x\times N_r\times N_\theta=100\times 40\times 160$ grid cells in axial, radial, and azimuthal direction required $10.2\,\text{h}$ user time for the simulation of $15$ advective time units running in parallel on $10$ CPU cores on a local cluster equipped with Intel\textsuperscript{\textregistered} Xeon\textsuperscript{\textregistered} E5-2630 ($2.40\,\mathrm{GHz}$) processors.
The corresponding standalone adaptive ODT simulation as discussed below required only $0.253\,\text{h}$ user time on a single core. 
This demonstrates the cost reduction achieved by reduced-order modeling with ODT while maintaining full-scale resolution.
This makes it feasible to simulate longer time series, higher Reynolds numbers, and a broader range of Prandtl numbers.

\subsection{\label{sec:vel} First-order velocity statistics}

\begin{figure}[t]
    \centering
        \includegraphics[scale=0.24]{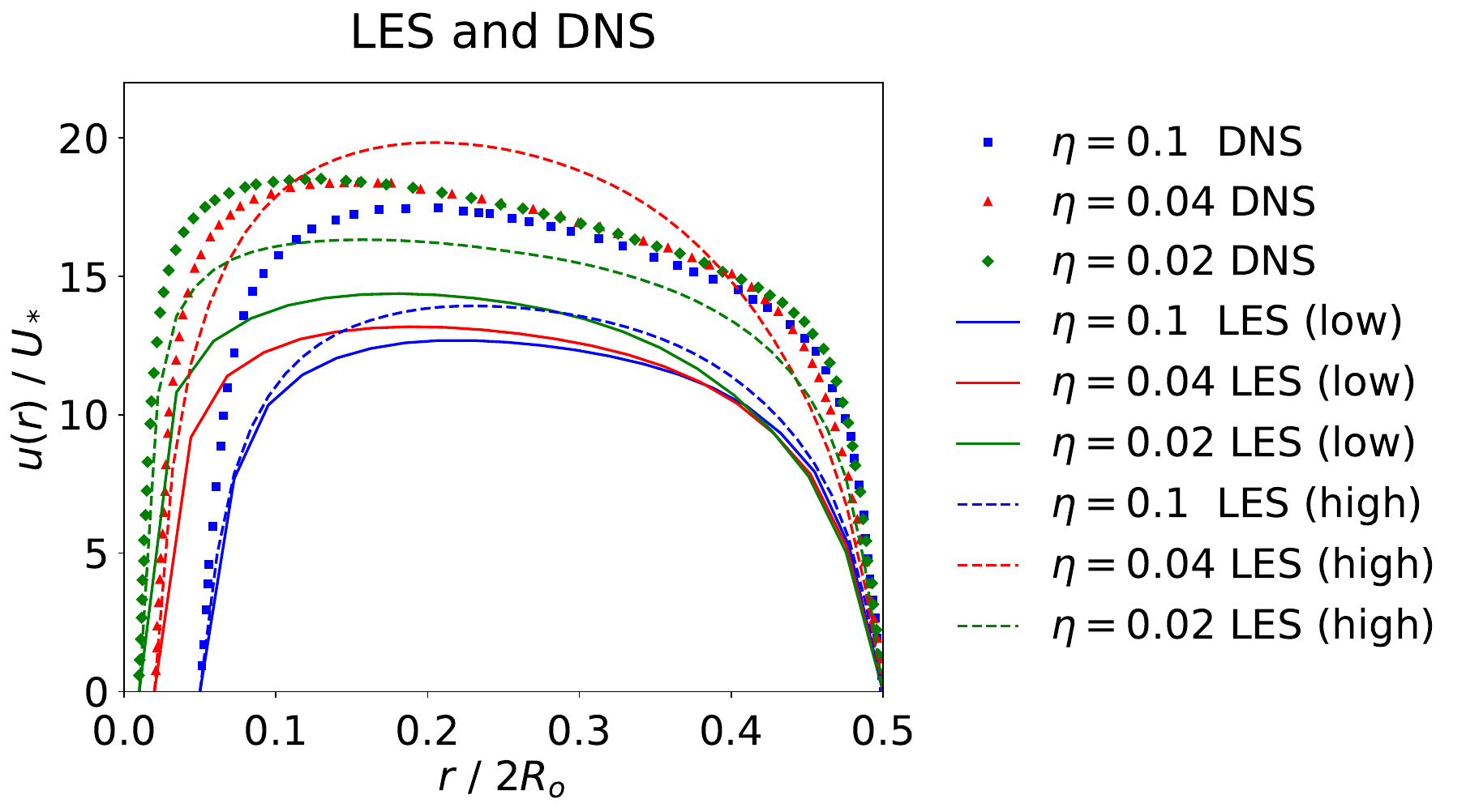}
        \includegraphics[scale=0.24]{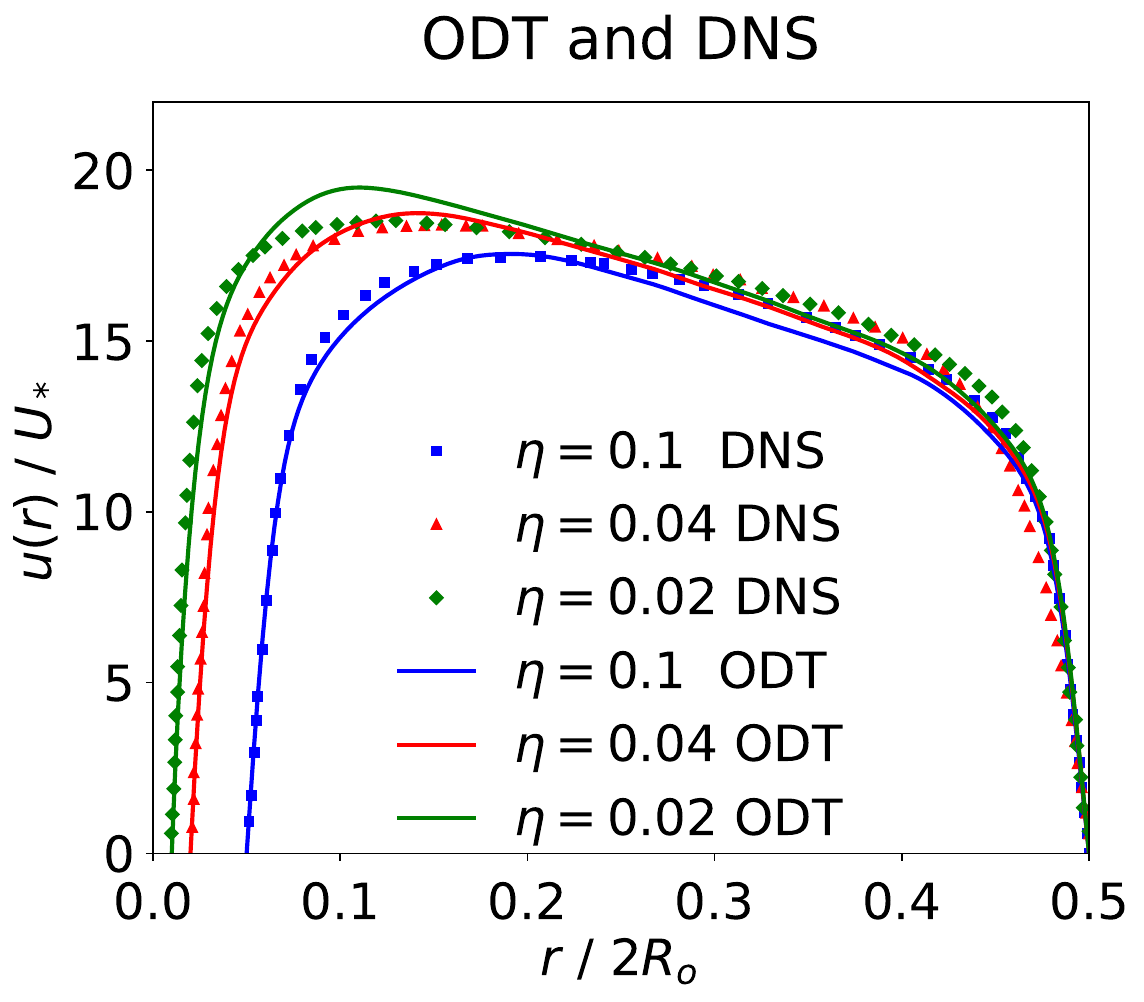}
    \caption{Mean velocity profiles $u(r)$ normalized by the friction velocity scale $U_*$ as function of the radial coordinate $r$.
    (\textit{left}) LES predictions and (\textit{right}) ODT predictions in comparison to reference DNS~\cite{boersma_etal:2011} results. 
    The equivalent circular pipe friction Reynolds number, $Re_* = 2U_* R_\text{o}/\nu = 600$, is kept fixed for all $\eta$.
    The label `low' indicates a coarse LES mesh, 
    while `high' indicates a finer mesh. 
    }
    \label{fig:Umean}
\end{figure}

Figure~\ref{fig:Umean} shows mean velocity profiles of concentric coaxial pipe flow predicted by a high resolution ($N_x\times N_r\times N_\theta = 100\times40\times160$) and low resolution ($50\times20\times80$) LES, as well as mesh-adaptive ODT simulations.
The LES (\textit{left}) is not capable of capturing the velocity profile.
The wall gradient is notably underestimated and the mean velocity may reach larger or smaller values than the DNS~\cite{boersma_etal:2011}.
In addition, present LES exhibit a strong grid dependence. 
These results suggest that radial transport processes over walls with spanwise curvature are poorly represented.
We attribute this to limitations of the wall model.
Albeit these may be circumvented by geometry-dependent parameterizations, present results demonstrate that LES can have limited predictive capabilities.

By contrast, ODT (\textit{right}), even as standalone model, does not exhibit such limitations.
The mean profile of the low-Reynolds number reference DNS~\cite{boersma_etal:2011} is very reasonably captured, in particular, in the vicinity of the wall.
Modeling errors manifest themselves primarily in the bulk, especially for the lowest radius ratio, $\eta=0.02$, investigated.
We attribute this behavior to unresolved three-dimensional flow structures around thin inner cylinders in addition to unresolved finite Reynolds number effects as the model prefers a fully developed turbulent state even at small Reynolds number.
Based on these results, we only consider ODT for further analysis of the thermal boundary layer discussed next.


\subsection{\label{sec:temp} First-order temperature statistics}

\begin{figure}[t]
    \centering
        \includegraphics[scale=0.24]{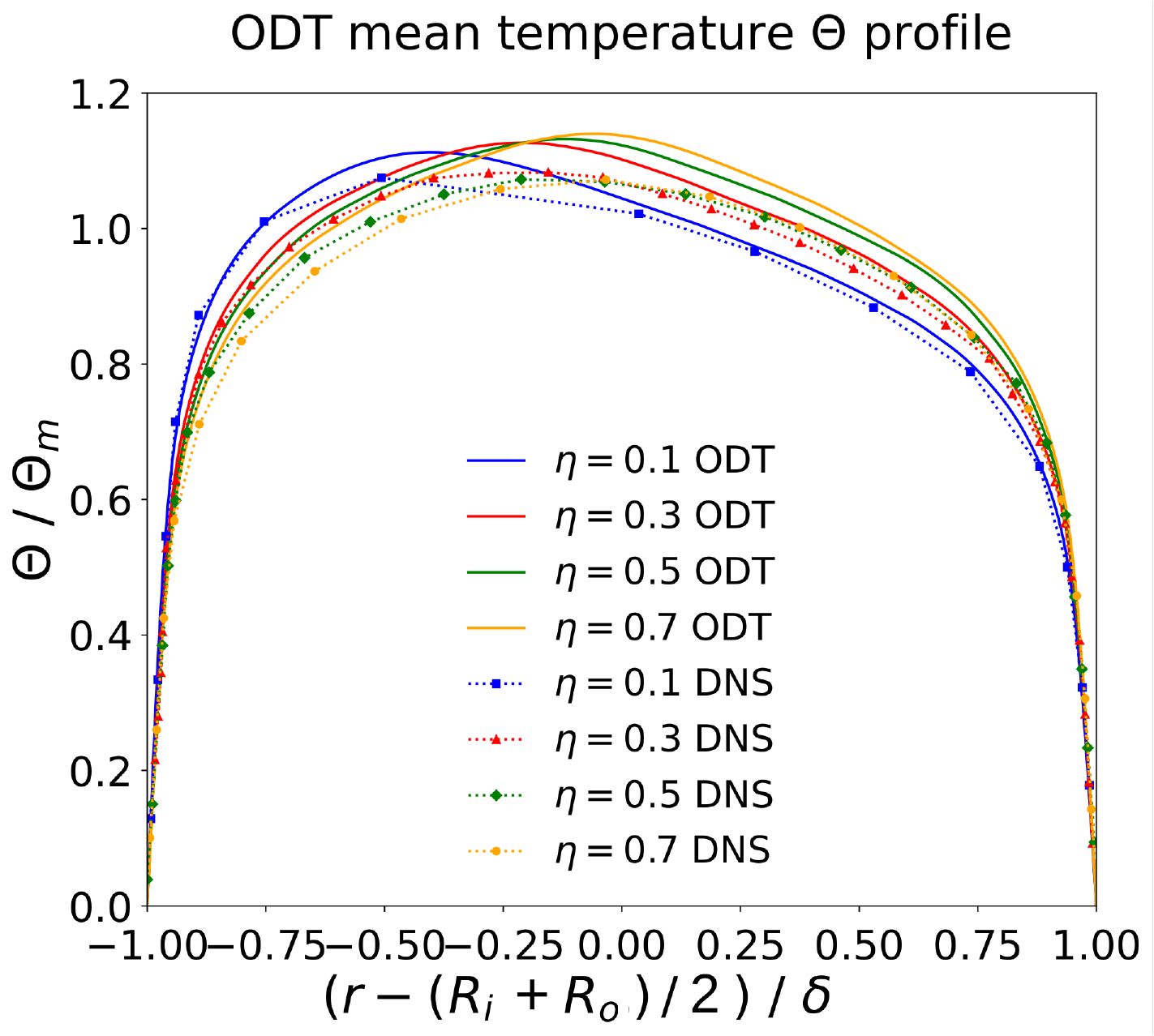}
        \includegraphics[scale=0.255]{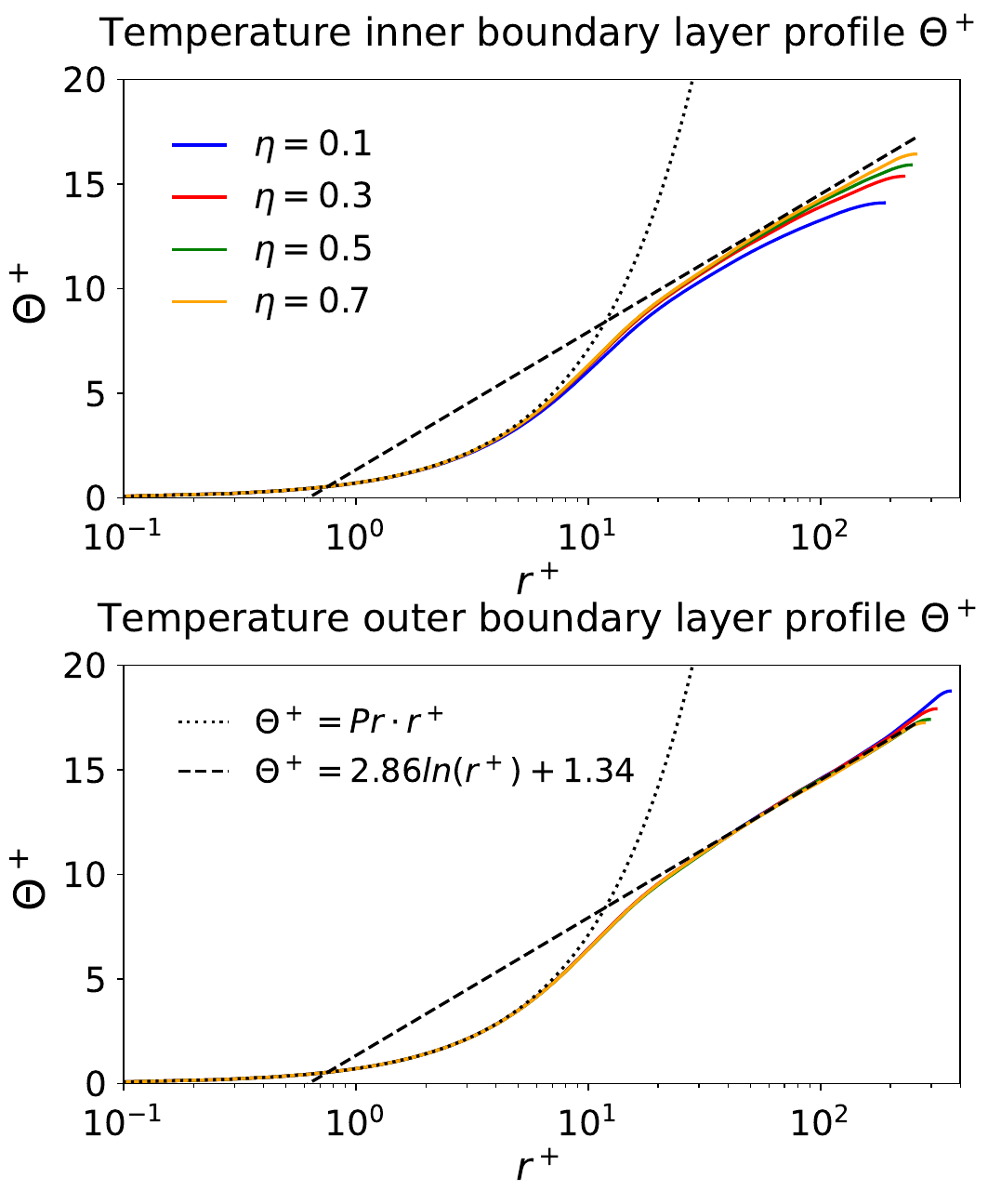}
    \caption{(\textit{left}) Mean perturbation temperature profiles $\Theta(r)$ for various radius ratios comparing ODT and reference DNS~\cite{bagheri_etal:2021} normalized by the bulk mean temperature $\Theta_\text{m}=2\int_{R_\text{i}}^{R_\text{o}} u\,\Theta\,r\,\text{d}r/(U_b\,(R_\text{o}^2-R_\text{i}^2))$, where $U_\text{b}=2\int_{R_\text{i}}^{R_\text{o}} u\,r\,\text{d}r/(R_\text{o}^2-R_\text{i}^2)$.
    The Prandtl, $Pr=0.71$, and bulk Reynolds number, $Re_\text{b}=2U_\text{b}\delta/\nu=17{,}700$ are kept fixed.
    (\textit{right}) Corresponding law-of-the-wall plots (\textit{top right}) over the inner cylinder and (\textit{bottom right}) over the outer cylinder for ODT only.}
    \label{fig:Tmean}
\end{figure}

Figure~\ref{fig:Tmean} (\textit{left}) shows radial profiles of the mean perturbation temperature for various radius ratios. 
It is observed that the ODT results reasonably reproduce the reference DNS \cite{bagheri_etal:2021}.
The wall gradients are accurately captured due to the calibration for scalar transfer \cite{Tsai_etal:2022}, but modeling errors manifest themselves by larger maximum values.
It is nevertheless remarkable that finite radius ratio effects, like the radial inner--outer asymmetry of the mean profiles, are reasonably captured by ODT as adjustable model parameters are kept fixed.
The thermal boundary layer over the outer wall is thicker than that over the inner wall due to curvature effects and flow features exhibiting smaller scales. 
As $\eta$ increases, the thickness of the thermal boundary layer over the inner wall increases nonlinearly in the same way as that over the outer wall decreases. 

Figure~\ref{fig:Tmean} (\textit{right}) shows law-of-the-wall plots of the thermal boundary layer (\textit{top right}) over the inner cylinder and (\textit{bottom right}) over the outer cylinder, respectively, for various $\eta$.
These profiles exhibit the linear conductive sublayer, $\Theta^+(r^+)=Pr\,r^+$, for $r^+\lesssim5$ and a turbulent logarithmic region for $r^+\gtrsim30$.  
The latter is parameterized as $\Theta^+(r^+)=\kappa_\Theta^{-1}\,\ln r^+ + B_\Theta$, where $\kappa_\Theta=0.35$ is the thermal von~K\'{a}rm\'{a}n constant and $B_\Theta=1.34$ the thermal additive constant that describe the weakly turbulent DNS~\cite{bagheri_etal:2021} data.
For the law-of-the-wall plots, reference DNS results are omitted in order to aid visual clarity. 
While present ODT data collapses well over the outer cylinder demonstrating inner layer similarity, wall-curvature effects prevent such a collapse for the inner cylinder. 
This is not a modeling error, but a correct representation of flow physics.
The origin of this effect is the nonlinear dependence of the boundary layer thicknesses on $\eta$ as noted above, which is a consequence of the generation of increasingly smaller scales in the flow as the inner cylinder curvature radius decreases.
By approximate analogy of scalar and axial momentum momentum transport towards the wall, the model results for the thermal boundary layer are, therefore, consistent with the argumentation put forward in \cite{boersma_etal:2011} for the momentum boundary layer.
This demonstrates that the fidelity and predictive capabilities of the ODT model are a result of the physics-based simultaneous resolution of heat and momentum transport on all relevant scales of the flow. 

\subsection{\label{sec:tflux} Second-order temperature statistics}

\begin{figure}[t]
    \centering
        \includegraphics[scale=0.205]{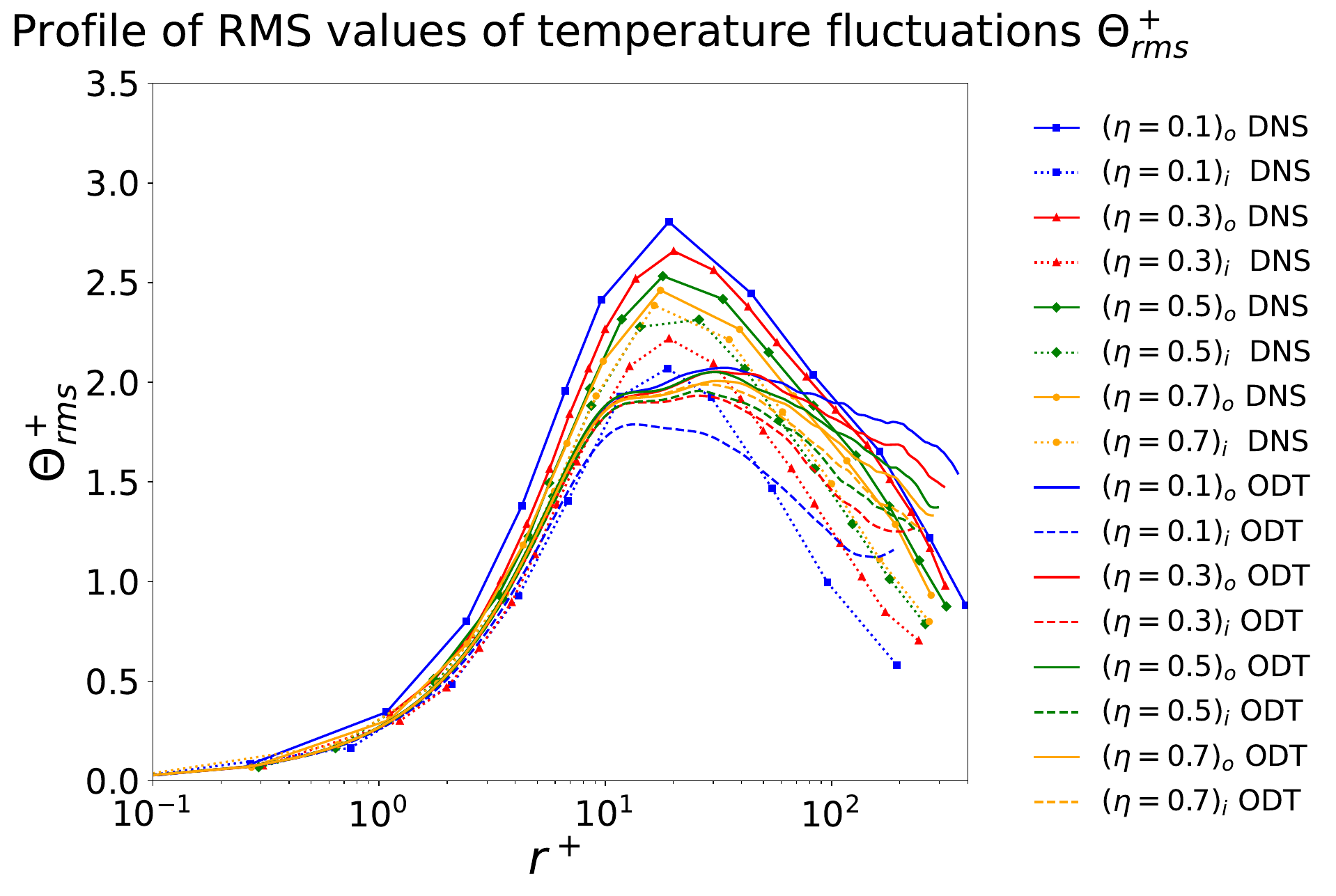}
        \includegraphics[scale=0.21]{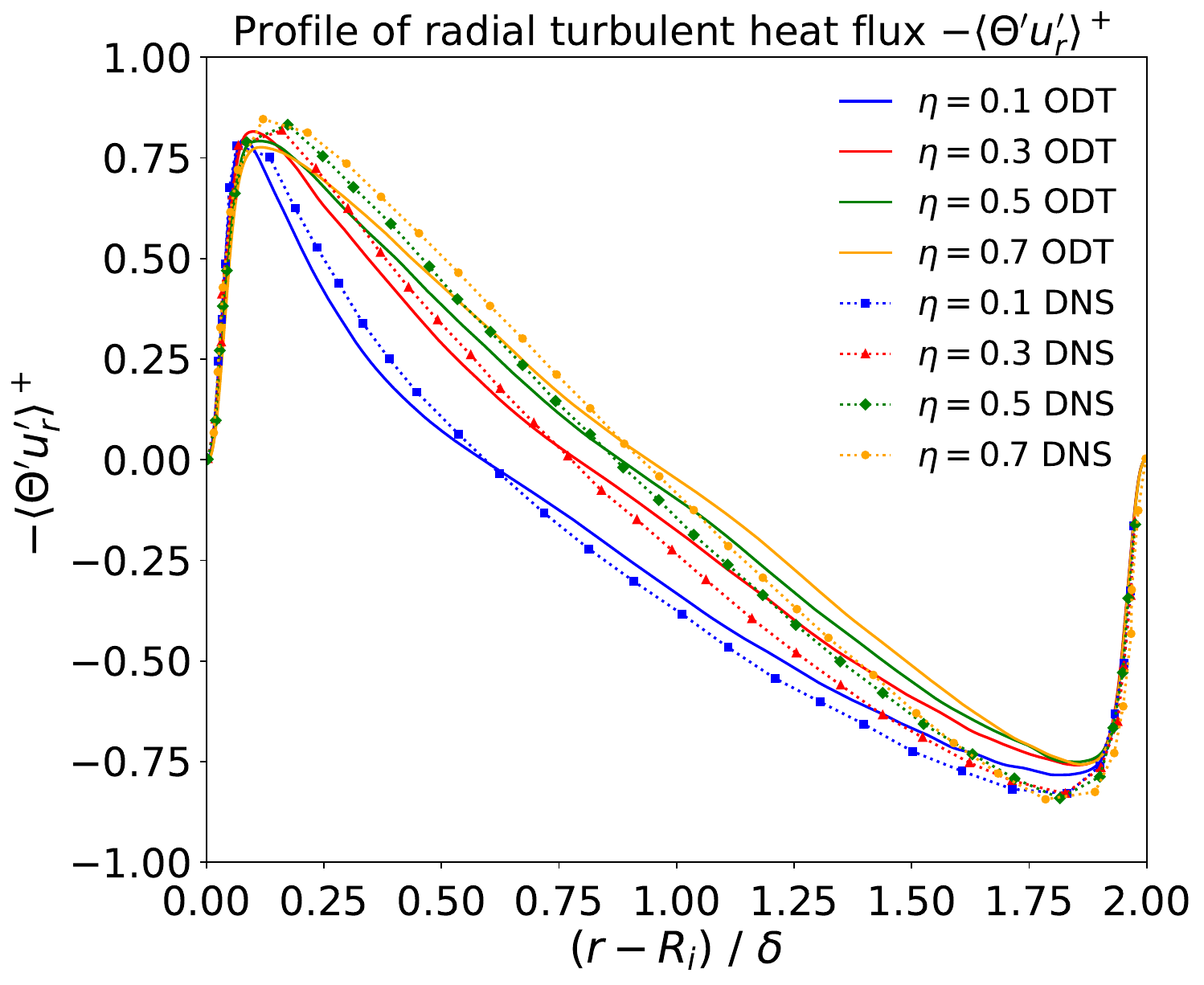}
    \caption{(\textit{left}) Boundary layer profiles of the normalized root-mean-square fluctuation temperature $\Theta^+_\text{rms}$ for various radius ratios showing ODT and reference DNS~\cite{bagheri_etal:2021} for fixed $Re_\text{b}=17{,}700$ and $Pr=0.71$.
    (\textit{right}) Corresponding bulk profiles of the turbulent radial heat flux $\langle \Theta^\prime u_{r}^\prime \rangle^+$ across the radial gap.}
    \label{fig:heatflux}
\end{figure}

Figure~\ref{fig:heatflux} (\textit{left}) shows the normalized root-mean-square (rms) fluctuation temperature $\Theta^+_\text{rms}$ various radius ratios comparing ODT with reference DNS \cite{bagheri_etal:2021}.
ODT significantly underestimates the near-wall turbulent fluctuation peak at $r^+\approx15$, but captures the trend with respect to $\eta$ with respect fluctuation magnitude and asymmetry between the boundary layer over the inner compared to the outer cylinder. 
The latter is consistent with the mean profiles discussed above, whereas the former is a known modeling artifact \cite{Lignell_etal:2013,Klein_etal:2022} that manifests itself also in pipe flows \cite{Lignell_etal:2018,Medina_etal:2019}.
The deficit in fluctuation magnitude does not necessarily imply incorrect radial fluxes.
On the contrary, capturing the radial asymmetry as noted above requires a physically correct representation of radial fluxes within the dimensionally reduced model.
This is made possible by the map-based turbulence modeling within the eddy events that take the role of advecting radial velocity fluctuations formally denoted by $u'_r$ below.

Figure~\ref{fig:heatflux} (\textit{right}) shows radial profiles of the normalized turbulent heat flux, $\langle \Theta^\prime u^\prime_r \rangle^+$, for various radius ratios.
Finite radius ratio effects dominate the spatial structure of $\langle \Theta^\prime u^\prime_r \rangle^+$ since wall-curvature effects govern the radial asymmetry of the flow profiles.
These geometrical features are reasonably captured by ODT.
This is revealed, in particular, by the zero-crossing $r_0$, where $\langle \Theta^\prime u^\prime_r \rangle^+=0$.
The radial location $r_0$ is insensitive to the Reynolds number, which suggests asymptotic turbulence properties in low Reynolds number ODT simulations.
Since $r_0$ corresponds with the location where $\Theta(r)$ reaches its maximum, a consistent representation of spanwise wall-curvature effects and wall-normal radial transport is therefore established.
The agreement between ODT and DNS is not perfect for the low Reynolds number investigated.
ODT is built for highly turbulent flows but no coaxial pipe DNS is available for this regime.
We therefore accept finite Reynolds number effects that manifest themselves by a slightly weaker turbulent heat flux in the bulk in comparison to reference DNS~\cite{bagheri_etal:2021}.
Weak turbulence only exhibits a small range of scales so that the ODT triplet mappings cannot carry the entire radial heat flux, inducing molecular diffusive fluxes.
The effect is weak and the modeling error vanishes for high asymptotic Reynolds numbers \cite{Klein_etal:2022}.
Based on the current model validation for concentric coaxial pipe flow, forthcoming ODT applications can be utilized for investigating scalar transfer at much higher Reynolds numbers.


\section{\label{sec:conc} Concluding remarks}

An advanced stochastic modeling strategy based on one-dimensional turbulence (ODT) has been utilized for accurate representation of fluctuating heat and momentum transfer processes in high Reynolds number concentric coaxial pipe flows.
It was shown that the standalone ODT model is capable of accurately capturing the radius ratio dependence of the mean flow, the thermal and momentum boundary layer, and the wall-normal (radial) turbulent heat flux. 
Modeling errors manifest themselves primarily by a lack in fluctuation variance at finite wall distance in the buffer layer and misrepresented turbulent fluxes at low Reynolds numbers at the edge of the applicability of the model. 
It was demonstrated that ODT overcomes present limitations in the predictive capabilities of a wall-modeled large-eddy simulation (WM-LES), which exhibits a strong grid dependence and only poorly captures the bulk and boundary layer flow.
ODT, therefore, offers new opportunities for high-fidelity modeling in process, thermal, and chemical engineering applications.

\section*{Acknowledgements}

This research is supported by the German Federal Government, the Federal Ministry of Education and Research and the State of Brandenburg within the framework of the joint project EIZ: Energy Innovation Center (project numbers 85056897 and 03SF0693A) with funds from the Structural Development Act (Strukturstärkungsgesetz) for coal-mining regions. M.K. acknowledges support by the BTU Graduate Research School (Conference Travel Grant). All authors thank Yousef Baroudi for initial LES tests within an EUNICE European University student exchange internship.

%
%


\bibliographystyle{spmpsci} 


\end{document}